\documentclass[showpacs,twocolumn,prl,aps,english,superscriptaddress]{revtex4}
\usepackage{amsmath}
\usepackage{graphicx}
\usepackage{amssymb}
\usepackage{babel}
\usepackage{color}
\usepackage{hyperref}
\usepackage{relsize}

\renewcommand{\v}[1]{|{#1}\rangle}
\newcommand{\bv}[1]{\langle{#1}|}

\newcommand{\A}{{^3\!\!A_2}}
\newcommand{\E}{{^3\!E}}
\newcommand{\At}{{^1\!A_1}}
\newcommand{\Et}{{^1\!E}}

\newcommand{\de}{{\delta\varepsilon}}

\newcommand{\I}{{\rm I}}
\newcommand{\II}{{\rm II}}
\newcommand{\LI}[2]{{\Lambda}^{(#1)}_{\I,#2}}
\newcommand{\LpI}[2]{{\bar\Lambda}^{(#1)}_{\I,#2}}
\newcommand{\LII}[2]{{\Lambda}^{(#1)}_{\II,#2}}

\newcommand{\lI}[1]{{\Lambda}^{(#1)}_{\I}}
\newcommand{\lpI}[1]{{\bar\Lambda}^{(#1)}_{\I}}
\newcommand{\lII}[1]{{\Lambda}^{(#1)}_{\II}}

\newcommand{\N}{{\cal N}}

\begin{document}

\author{Dmitry Solenov}
\email{d.solenov@gmail.com}
\altaffiliation{Present address: Naval Research Laboratory, 4555 Overlook Ave., SW Washington, District of Columbia 20375, USA}
\affiliation{National Research Council, National Academies, Washington, District of Columbia 20001, USA}

\author{Sophia E. Economou}
\affiliation{Naval Research Laboratory, Washington, District of
Columbia 20375, USA}
\author{Thomas L. Reinecke}
\affiliation{Naval Research Laboratory, Washington, District of
Columbia 20375, USA}

\title{Two-qubit quantum gates for defect qubits in diamond and similar systems}

\begin{abstract}
We propose a fast, scalable all-optical design for arbitrary two-qubit operations for defect qubits in diamond (NV centers) and in silicon carbide, which are promising candidates for room temperature quantum computing. The interaction between qubits is carried out by microcavity photons. The approach uses constructive interference from higher energy excited states activated by optical control.  In this approach the cavity mode remains off-resonance with the directly accessible optical transitions used for initialization and readout. All quantum operations are controlled by near-resonant narrow-bandwidth optical pulses. We perform full quantum numerical modeling of the proposed gates and show that high-fidelity operations can be obtained with realistic parameters.
\end{abstract}


\maketitle

Qubits encoded by electron states of defects in diamond and in silicon carbide have become promising candidates for room temperature quantum information processing. Long coherence times of $\sim 50-200 \mu$s, initialization, readout and single qubit operations have been demonstrated \cite{Kennedy,Awschalom-SiC} for qubits in both systems at room temperature. Simple entangling operations have been demonstrated in the system of a negatively charged nitrogen vacancy center (NV) in diamond coupled to a nearest carbon (C) \cite{Jelezko,Dutt} or a nearest nitrogen (N) \cite{Awschalom-NVC-N,Gaebel-NVC-N} nuclear spin. However these systems are not scalable. 
Entangling operations of two distant NV qubits have recently been performed in a challenging experiment by joint measurement of the photons emitted by the two NV centers \cite{Bernien}. The challenges associated with using this type of entanglement in a quantum information processor impose stringent requirements on the frequencies of the NV centers. Further, this approach is probabilistic---only $\sim3\%$ of the emitted
photons come from the zero phonon line of the NV center \cite{Bernien}---resulting in a low rate of successful operations.
It has also been proposed that nitrogen defect nuclear spins in diamond might potentially be used to mediate a long-range interaction between distant NV qubits \cite{Awschalom-NVC-N}. However, introduction of a large number of defect nuclear spins, as well as problems caused by imprecise defect positioning in such systems, makes this impractical. Thus, experimentally viable deterministic (reversible) two-qubit gates between distant defect qubits remain an important challenge.

Recent developments in photonic microcavities in diamond \cite{diamondCav-ring,diamondCav-ph-crystal} and in silicon carbide \cite{SiC-cav-Yamada} have opened opportunities to couple distant defect qubits via cavity photons.  Microcavities in these systems have been fabricated in both ring \cite{diamondCav-ring} and void-pillar geometries \cite{diamondCav-ph-crystal} with defect qubits placed near the surface. High quality factors of optical modes and large values of photon coupling to the defect optical transitions have been reported \cite{SiC-cav-Yamada,diamondCav-ring,Faraon}. This has made photon-mediated distant qubit-qubit interactions possible experimentally.

We propose a robust all-optical approach to deterministic two-qubit quantum operations for NV-center qubits in diamond and similar defect qubits in other systems, including silicon carbide. The interaction between defect qubits is carried out by a photonic microcavity mode, that remains off-resonance with dipole optical transitions involving low energy excitations. This leads to effective isolation of the single-defect excitations in each qubit. As a result, a $\Lambda$-system involving the lowest optically accessible excited states in each defect can be used to carry out single-qubit operations, readout, or initialize qubits to a specific single-qubit state optically. The higher-energy states involving excitations in both defects do not decouple from the cavity as effectively and can mediate interactions between the qubits. Although these non-local states are not directly accessible via selective narrow-band pulses, an optical activation can be performed to gain spectrally selective pulse control needed to fully manipulate the two-qubit system. We introduce a framework that can be used to perform an arbitrary two-qubit operation with only a few pulses. Particular attention is paid to conditional (control) two qubit operations, $C(U)$, that are needed for most quantum algorithms. The approach developed in this work provides direct access to a variety of fast two-qubit operations and has potential for scalability. We give detailed examples for several optically-controlled two-qubit gates including control-Z (CZ), control-NOT (CNOT), two-qubit swap (SWAP), and control-phase [$C(\phi)$]. We perform full quantum mechanical simulations of the system at low temperatures and find that high values of fidelity for two-qubit operations can be achieved in the strong coupling regime (low cavity losses) for experimentally reasonable parameters \cite{Faraon,Lukin-Talk, Faraon2}. 

%====================================================================
\begin{figure}
\centering{}
\includegraphics[width=0.8\columnwidth]{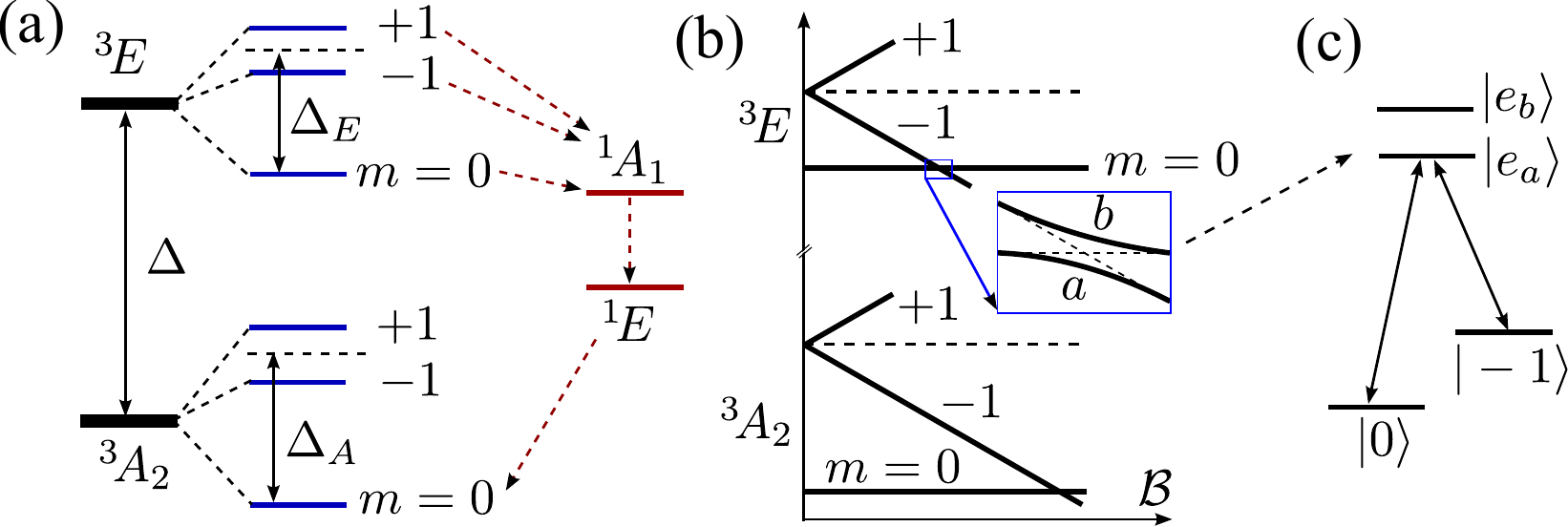}
\caption{\label{fig1}
Nitrogen-vacancy center (NV) in diamond as a qubit.
(a) The relevant states of the NV in diamond \cite{Acosta}. The qubit is encoded by states $\A_{m=0}$ and $\A_{m=-1}$. Non-radiative decay of excited states takes place via $^1\!E$ and $^1\!A_1$ (red) \cite{Acosta, Doherty}. (b) Energies of one NV center as a function of magnetic field. (c) A $\Lambda$-system accessible due to mixing between states $\v{0}$ and $\v{-1}$ in $\E$ triplet.
}
\end{figure}
%====================================================================
%====================================================================
\begin{figure}
\centering{}
\includegraphics[width=1.0\columnwidth]{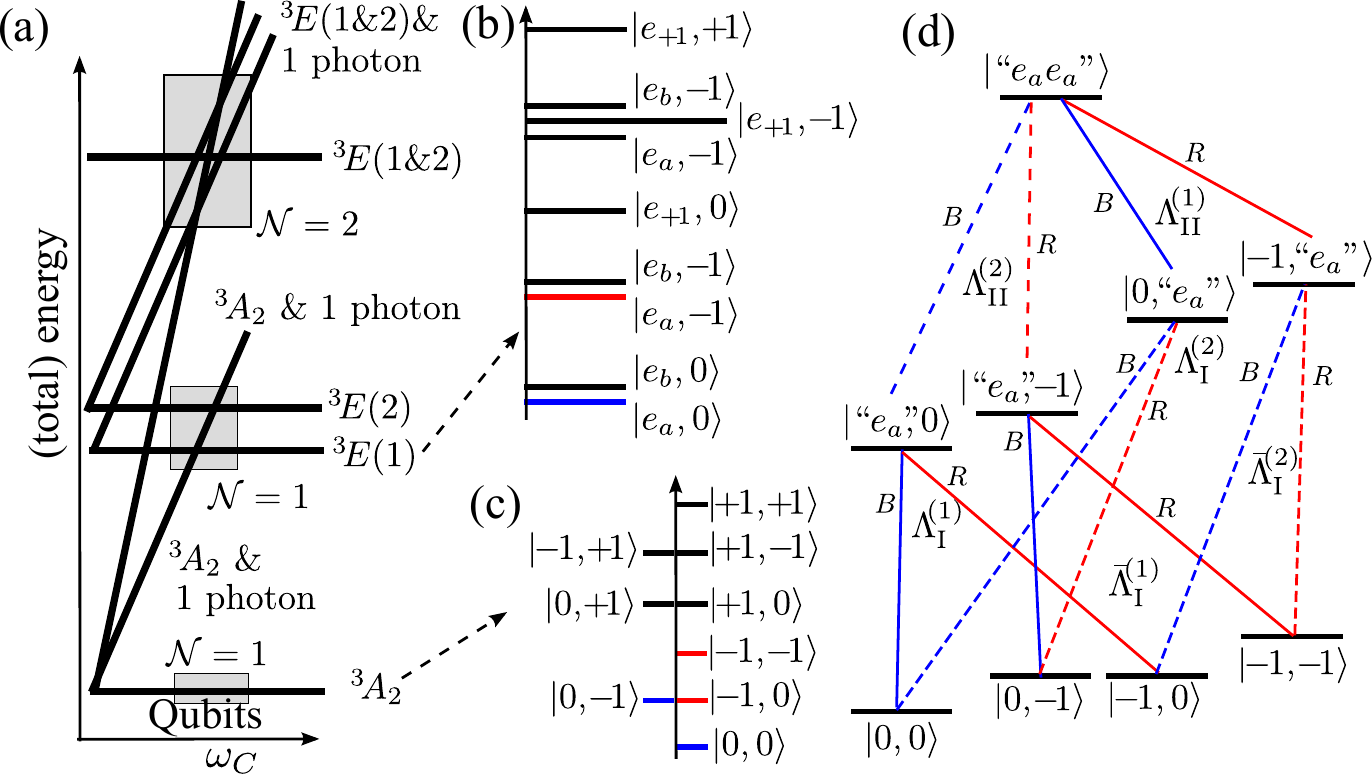}
\caption{\label{fig2}
The energies of the defect-cavity system and $\Lambda$-system for single- and two-qubit manipulations.
(a) Schematic depiction of the spectrum of two defects in a cavity (total energy) for varying values of $\omega_C$. The spectrum can be classified according to the total number of optical excitations $\N$. The $\N=0$ part, shown in panel (c), is used to encode the qubits. Panel (b) shows the $\N=1$ part (schematically).
(d) Accessible $\Lambda$-systems involving $\v{``e_a"}$ states. A similar hierarchy can be constructed using $\v{``e_a"}$ states. The higher and lower (transition) frequency leg of each $\Lambda$-system is denoted by B and R, respectively.
}
\end{figure}
%====================================================================

For definiteness we will focus mainly on the case of NV centers in diamond. Other related systems, such as defect centers in silicon carbide \cite{Acosta,Baranova,Gali,Riedel}, are similar. The NV defect has eight states of interest \cite{Acosta,MazeLukin} [see Fig.~\ref{fig1}(a)], six of which, $\A_m$ and $\E_m$ ($m=0,\pm 1$), participate in optical transitions. The other two states, $\v{\At}$ and $\v{\Et}$, are involved in non-radiative recombination from $\E_m$ \cite{Acosta} and are important for initialization and measurement \cite{Acosta,Awschalom-SiC,Kennedy,Awschalom-NVC-N}. The two ($\v{0}$ and $\v{-1}$) of the three lowest spin-1 states, $\A_m$, are used to encode the qubit. An external magnetic field, ${\cal B}$, splits the two triplet states as 
$H_g = \sum_m(|m|\Delta_A\! +\! m g {\cal B})\v{m}\bv{m}$ 
and 
$H_e = \sum_m(\Delta\! +\! |m|\Delta_E\! +\! m g {\cal B})\v{e_m}\bv{e_m}$ 
for $\A_m$ and $\E_m$ triplets respectively [see Fig.~\ref{fig1}(b)]. The optical transition energy $\Delta$ depends on the local environment of each defect, particularly when the defect is on the surface of a microcavity \cite{Batalov,Bassett,Riedel}, and can vary substantially from defect to defect. Typical entangling gate schemes \cite{nielsenchuang,Nori} operate by bringing the states into and out of resonance with the cavity, thus, switching the interaction on and off to perform the gate. Our approach does not use such control and, hence, does not require resonance between cavity photons and dipole transition lines in each defect. Instead, the natural variations in $\Delta$ are used to isolate each defect spectrally. Isolated defect qubits in diamond and silicon carbide systems are typically manipulated with microwave pulses \cite{Kennedy,Awschalom-SiC,Awschalom-NVC-N,Gaebel-NVC-N}. Recently, however, mixing between $\E$ triplet states, $\v{e_m}$ [given by $H'_e = g_0(\v{e_0}\bv{e_{-\!1}}+h.c.)$],
has been used to optically manipulate single-qubit states at non-zero magnetic field \cite{Togan,Yale}. 
The coupling $g_0$ arises due to local crystal strains and non-uniform electrostatic charges that break the $C_{3v}$ symmetry of the NV defect \cite{Bassett}. It results in an avoided crossing between states $\v{e_0}$ and $\v{e_{-1}}$ at some value of magnetic field [see Fig.~\ref{fig1}(b)] and provides a $\Lambda$-system [see Fig.~\ref{fig1}(c)] for optical manipulations. The total Hamiltonian of an NV defect system is
\begin{equation}\label{eq:H-NV}
H_{NV} = H_g + H_e + H'_e + H_{NR}.
\end{equation}
The last term describes $\At$ and $\Et$ states, as well as interaction with phonons. It will be omitted in the discussion of the coherent quantum gates. At the same time, states $\At$ and $\Et$ are important and will contribute to the evolution of the non-coherent system investigated in our numerical simulations. They provide the dominant (non-radiative) decay pathway that affects the fidelity of all two- and single-qubit optical manipulations.

The proposed framework for an arbitrary two-qubit operation is based on a hierarchy of local (single-defect) and non-local (two defects) $\Lambda$-systems. In order to construct it we first analyze the spectrum of two defects and a microcavity mode. In the rotating wave approximation, the spectrum is given by the Hamiltonian
\begin{equation}\label{eq:H0}
H_0 = \sum_{i=1,2}\left[ H_{NV}^{(i)}+H_I^{(i)} \right] + \omega_C a^\dag a,
\end{equation}
where $H_I=\gamma\sum_m(\v{m}\bv{e_m}\,a^\dag + h.c.)$ for each defect and $\delta\Delta = \Delta^{(2)}-\Delta^{(1)}>0$ (the superscripts refer to defect 1 and 2). The Hamiltonian (\ref{eq:H0}) conserves the total number of excitations 
$\N = n^{(1)} + n^{(2)} + a^\dag a$ (where for each defect $n=\sum_m\v{e_m}\bv{e_m}$). The general structure of the states as functions of the cavity mode frequency $\omega_C$ is shown in Fig.~\ref{fig2}(a). The $\N=0$ subspace is unaffected by the cavity [Fig.~\ref{fig2}(c)] and contains the qubit states. In the $\N=1$ subspace, the cavity shifts energies and mixes states $\E_m$ and $\A_m$. For practical values of $\delta\Delta$ this change is local, similar to the case of a dot-cavity system in Ref.~\cite{solenov-QDs}. The three $\N=1$ bands in the total energy spectrum correspond to one excitation in one of the two defects or in the cavity. The lowest four states are $\v{``e_a",0}$, $\v{``e_b",0}$, $\v{``e_a",-1}$, and $\v{``e_b",-1}$ [see Figs.~\ref{fig1}(c) and \ref{fig2}(b)]. The quotation marks indicate that the states are mixed (locally) with single photon states. For non-zero values of $g_0$, dipole transitions involving $\N=1$ as the transition state can be grouped into several $\Lambda$-systems [see also Fig.~\ref{fig2}(d)]: $\lI{1}\to$\{$\v{0,0}$, $\v{``e_a",0}$, $\v{-1,0}$\}, $\lpI{1}\to$ \{$\v{0,-1}$, $\v{``e_a",-1}$, $\v{-1,-1}$\}, $\lI{2}\to$ \{$\v{0,0}$, $\v{0,``e_a"}$, $\v{-1,0}$\}, and $\lpI{2}\to$ \{$\v{0,-1}$, $\v{-1,``e_a"}$, $\v{-1,-1}$\}. Transitions reaching to states with $\N=2$ are grouped into $\lII{1}\to$ \{$\v{0,``e_a"}$, $\v{``e_ae_a"}$, $\v{-1,``e_a"}$\}, $\lII{2}\to$ \{$\v{``e_a",0}$, $\v{``e_ae_a"}$, $\v{``e_a",-1}$\}. For clarity of presentation we will omit transitions involving $e_b$, which are similar. In the basis \{$\v{0,0}$, $\v{0,-1}$, $\v{-1,0}$, $\v{-1,-1}$, $\v{``e_a",0}$, $\v{``e_a",-1}$, $\v{0,``e_a"}$, $\v{-1,``e_a"}$, $\v{``e_ae_a"}$\} the reduced Hamiltonian with control pulses takes the form
\begin{equation}\label{eq:Pulses}
{\tiny
\left(
\begin{array}{cccc|cccc|c}
%	00		01		10		11		e0		e1		0e		1e		ee
	0&		0&		0&		0&		\LI{1}{B}&	0&		\LI{2}{B}&	0&		0\\% 00
	0&		E&		0&		0&		0&		\LpI{1}{B}&	\LI{2}{R}&	0&		0\\% 01
	0&		0&		E&		0&		\LI{1}{R}&	0&		0&		\LpI{2}{B}&	0\\% 10
	0&		0&		0&		2E&		0&		\LpI{1}{R}&	0&		\LpI{2}{R}&	0\\% 11
\hline
	\LI{1}{B}&	0&		\LI{1}{R}&	0&		E^1_e&		0&		0&		0&		\LII{2}{B}\\% e0
	0&		\LpI{1}{B}&	0&		\LpI{1}{R}&	0&		\bar E^1_e\!\!+\!E&	0&		0&		\LII{2}{R}\\% e1
	\LI{2}{B}&	\LI{2}{R}&	0&		0&		0&		0&		E^2_e&		0&		\LII{1}{B}\\% 0e
	0&		0&		\LpI{2}{B}&	\LpI{2}{R}&	0&		0&		0&		\bar E^2_e\!\!+\!E&	\LII{1}{R}\\% 1e
\hline
	0&		0&		0&		0&		\LII{2}{B}&	\LII{2}{R}&	\LII{1}{B}&	\LII{1}{R}&	E^1_e\!\!+\!E^2_e\!\!-\!\de\\% ee
\end{array}
\right)\!\!\!,
}
\end{equation}
where $E=\Delta_A-gB$, $E^n_e=\Delta+\Delta_E-gB-\de_n + \delta_{2,n}\delta\Delta$, $\bar E^n_e=\Delta+\Delta_E-gB-\bar\de_n + \delta_{2,n}\delta\Delta$. Factors $e^{i\omega t}$ with corresponding (near-resonant) pulse frequencies $\omega$ are not shown to shorten notation. Unless $\gamma/\delta\Delta\sim 1$, the shifts due to the cavity $\de_n$ and $\bar\de_n$ are not distinguishable from each other as compared to the cavity-induced shift $\de$ of the $\N=2$ state, i.e. $|\de_n-\bar\de_n|\ll\de$. For a typical regime of $\gamma/\delta\Delta\ll 1$, the difference $\de_n-\bar\de_n$ and $\de$ are proportional to $1/\delta\Delta^2$ and $1/\delta\Delta$ respectively \cite{solenov-QDs}). Thus, we can use $E^n_e=\bar E^n_e$, $\lI{n}=\lpI{n}$. As a result, each $\lI{n}$, when addressed individually, operates only on $n$-th qubit and provides a full set of single-qubit manipulations \cite{Yale,economouprb1}.

Note that single-qubit operations \cite{Yale,economouprb,rosenzener} performed via $\lI{n}$ cannot be done concurrently. Moreover if the system is not returned to the qubit subspace after a set of, e.g., $\lI{1}$ pulses, subsequent ``single-qubit" operations induced by, e.g., $\lI{2}$ pulses become non-local. The non-locality is rooted in the cavity-induced interaction, and comes from $\de\neq 0$, which, in this example, makes $\lI{2}$ and $\lII{2}$ spectrally distinct. One of the important classes of two-qubit operations that is a consequence of such non-commutativity of single-qubit controls is $C(U)$ operations,
\begin{equation}\label{eq:CU}
C(U) = \left({\small
\begin{array}{cccc}
\begin{array}{cccc}
1 & 0\\
0 & 1\\
\end{array}
& 
\begin{array}{cccc}
0 & 0\\
0 & 0\\
\end{array}
\\
\begin{array}{cccc}
0 & 0\\
0 & 0\\
\end{array}
&
\mathlarger{\mathlarger{\mathlarger{\mathlarger{U}}}}
\\
\end{array}
}\right),
\end{equation}
where $U$ is an arbitrary unitary $2\times 2$ matrix. Using the $n$-th qubit as a control qubit and the $m$-th qubit as a target qubit, $C(U)$ is performed by a series of pulses \{$\{\LI{n}{i},2\pi M+\pi\}$, $[\lI{m}]$,$\{\LI{n}{i},2\pi M'+\pi\}$. Here $[\lI{m}]$ is a single two-color pulse or a series of simple pulses that would normally perform a single qubit operation $U$ on $m$-th qubit. For $C(U)$ shown in Eq.~(\ref{eq:CU}) $i=B$. When $i=R$, the matrix $U$ operates on the first two states, rather then the last two. The first and the last pulses are resonant $\pi$ (swap) pulses, with additional $2\pi M$ and $2\pi M'$ rotations that are used to correct for incurred single-qubit phases. 

Other two-qubit operations can also be performed in a similar manner. More generally, a single four or six-color (phase locked) pulse operating in $\Lambda_\I$ and/or $\Lambda_\II$ with appropriate shape can be used to perform an arbitrary two-qubit rotation. Finding optimal pulse shapes for such multi-color controls, however, must be addressed numerically. In Table~\ref{table:one} we list examples of the most frequently required two-qubit operations performed with a series of simple pulses. The CZ gates can also be performed in the case when mixing between $\E$ triplet states does not occur. However, an additional microwave control is needed to construct other two-qubit gates in this case.

\begin{table}
\centering\scriptsize
\caption{Examples of all-optical two-qubit gates. Each pulse is defined by the targeted transition [see Fig.~\ref{fig2}(d)], strength, and detuning (omitted when zero). It is assumed that pulses do not overlap appreciatively. Multiple options are available for each gate (only several are shown, as an example). Global phase is ignored. The last column shows states out of the qubit subspace that are involved during normal coherent  operation. Pulses do not have to be phase-locked, unless indicated otherwise.}
\begin{tabular}{|l|l|l|}
\hline
{\bf gate}&{\bf pulse sequence}&{\bf states\! involved}	\\ 
\hline 
CZ &&\\
\, {\rm diag}\{-1,1,1,1\} & \{\!\{$\LI{1}{R}$,$\pi$\},\{$\LI{2}{B}$,$2\pi$\},\{$\LI{1}{R}$,$\pi$\}\!\} & $\v{0,\!0}\!$,$\v{\!-\!1\!,\!0}\!$,$\v{\!-\!1\!,\!\!-\!1}\!$\\
\, {\rm diag}\{1,-1,1,1\} & \{\!\{$\LI{1}{B}$,$\pi$\},\{$\LII{2}{B}$,$2\pi$\},\{$\LI{1}{B}$,$\pi$\}\!\} & $\v{0,\!0}\!$,$\v{0,\!\!-\!1}\!$\\
			  & \{\!\{$\LI{2}{B}$,$\pi$\},\{$\LI{1}{B}$,$2\pi$\},\{$\LI{2}{B}$,$3\pi$\}\!\} & $\v{0,\!0}\!$,$\v{0,\!\!-\!1}\!$,$\v{\!-\!1,\!0}\!$\\
\, {\rm diag}\{1,1,-1,1\} & \{\!\{$\LI{1}{B}$,$\pi$\},\{$\LI{2}{B}$,$2\pi$\},\{$\LI{1}{B}$,$3\pi$\}\!\} & $\v{0,\!0}\!$,$\v{0,\!\!-\!1}\!$,$\v{\!-\!1,\!0}\!$\\
			  & \{\!\{$\LI{2}{B}$,$\pi$\},\{$\LII{1}{B}$,$2\pi$\},\{$\LI{2}{B}$,$\pi$\}\!\} & $\v{0,\!0}\!$,$\v{\!-\!1,0}\!$\\
\, {\rm diag}\{1,1,1,-1\} & \{\!\{$\LI{2}{B}$,$\pi$\},\{$\LI{1}{R}$,$2\pi$\},\{$\LI{2}{B}$,$\pi$\}\!\} & $\v{0,\!0}\!$,$\v{\!-\!1\!,\!0}\!$,$\v{\!-\!1\!,\!\!-\!1}\!$\\
\hline
CNOT &&\\
\, control: QB 1 & \{\!\{$\LI{1}{B}$,$\pi$\},  [\{$\LI{2}{B}$,$\pi$\},\{$\LI{2}{R}$,$\pi$\}, & all \\
		  &\{$\LI{2}{B}$,$\pi$\}]$^*$,  \{$\LI{1}{B}$,$\pi$\}\!\} &\\
\, control: QB 2 & \{\!\{$\LI{2}{B}$,$\pi$\},  [\{$\LI{1}{B}$,$\pi$\},\{$\LI{1}{R}$,$\pi$\}, & all \\
		  &\{$\LI{1}{B}$,$\pi$\}]$^*$,  \{$\LI{2}{B}$,$\pi$\}\!\} &\\
\hline 
SWAP 	&\{\!\{$\LI{1}{B}$,$\pi$\},\{$\LI{2}{B}$,$\pi$\}, [\{$\LII{2}{R}$,$\pi$\},	& $\v{0,\!0}\!$,$\v{0,\!\!-\!1}\!$,$\v{\!-\!1,\!0}\!$ \\
	&\{$\LII{1}{R}$,$\pi$\},\{$\LII{2}{R}$,$3\pi$\}]$^*\!\!\!$, \{$\LI{2}{B}$,$\pi$\},	&\\
	&\{$\LI{1}{B}$,$3\pi$\}\!\}	&\\
\hline 
C($\phi$) &&\\
\, {\rm diag}\{1,1,1,$e^{i\phi}$\} 	& \{\!\{$\LI{2}{B}$,$\pi$\}\!,\{$\LI{1}{R}$,$2\pi$,$\delta$\}\!,\{$\LI{2}{B}$,$3\pi$\}\!\} & $\v{0,\!0}\!$,$\v{\!-\!1\!,\!0}\!$,$\v{\!-\!1\!,\!\!-\!1}\!$\\
\hline 
\end{tabular}
{$^*$ Pulses have to be phase-locked; the sequence can also be performed via a single two-color off-resonant pulse.}
\label{table:one}
\end{table}

In a fully coherent system all proposed operations can be performed with fidelity $F=1$. As a result of decoherence processes, such as decay from the excited states of the defects and cavity leakage, a trade-off must be made in designing the gates. Shorter control pulses reduce dissipative losses, but result in more destructive interference by coupling to other optical transitions. Longer (smaller bandwidth) pulses cause less destructive interference but increase real losses. In diamond defect qubits and other similar systems, such as silicon carbide, the time scale of the qubit decoherence is large, $\sim \mu$s \cite{Kennedy,Awschalom-SiC}. However, spontaneous decay of excited states takes place at a much shorter time scale, $\sim 10$ ns \cite{Acosta} or less. In order to investigate coherence of quantum gates based on $\Lambda^\I$ and $\Lambda^\II$ we perform full quantum simulations of the entire system. We include all eight states shown in Fig.~\ref{fig1}(a) for each defect system (the first two states are the qubit states) and ten states to represent the microcavity mode. As a result of substantial non-linearity caused by adsorption and emission of cavity photons by defect states in the regime of interest, as well as the use of narrow-bandwidth control pulses, the microcavity mode population remains low.

\begin{figure}
\centering{}
\includegraphics[width=1.0\columnwidth]{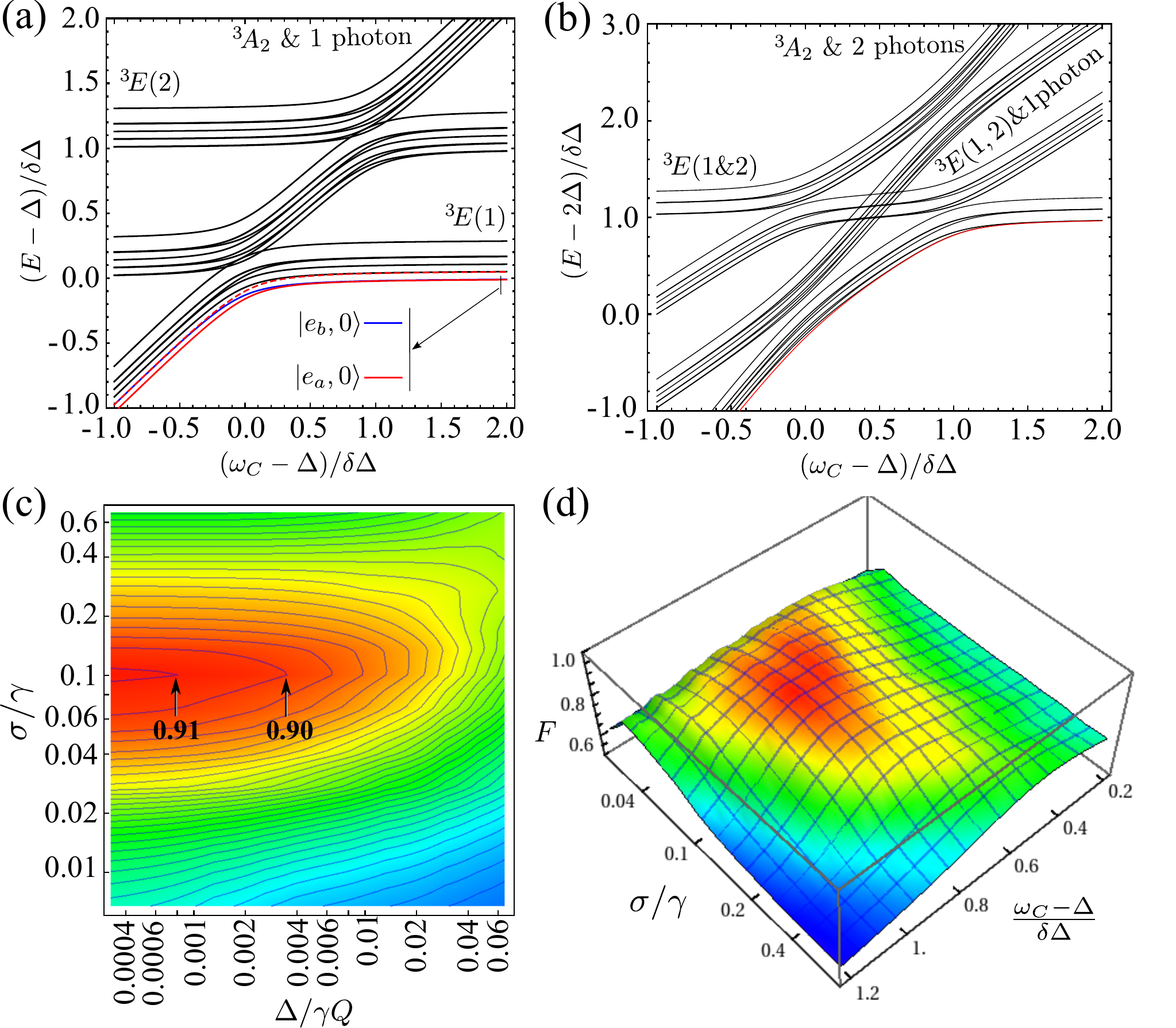}
\caption{\label{fig3}
Fidelity of the CZ gate. Panels (a) and (b) show $\N=1$ and $\N=2$ parts of the spectrum respectively. (c) Average fidelity as a function of pulse bandwidth, $\sigma$, and the cavity quality factor $Q$, optimized over positioning of $\omega_C$. (d) Fidelity as a function of $\sigma$ and $\omega_C$ for the leftmost value of $\Delta/\gamma Q$ in panel (c).
}
\end{figure}

As an example, we chose a diamond NV system with $\gamma=15$ $\mu$eV, $\delta\Delta=100$ $\mu$eV, and $g_0=0.1$ $\mu$eV. The energy spectrum (total energy) of the system near the cavity-defect anti-crossings in subspace ${\cal N}=1$ and ${\cal N}=2$ is shown in Fig.~\ref{fig3}(a) and (b) respectively. The top state $\v{``e_a",0}$ of the $\lI{1}$ system is the lowest energy state in Fig.~\ref{fig3}(a). Note that the energy difference between  $\v{``e_a",-1}$ and $\v{``e_a",0}$ [the third and the first lowest states in Fig.~\ref{fig3}(a) respectively] is the same for any microcavity frequency. This makes $\lI{1}$ and $\lpI{1}$ practically identical, as discussed earlier. The first state in Fig.~\ref{fig3}(b), $\v{``e_ae_a"}$, is a part of the $\lII{n}$ system. The energy of this state deviates significantly from the sum of energies of $\v{``e_a",0}$ and $\v{0,``e_a"}$ near the anti-crossings with the photon band, and thus $\de\neq 0$.

%====================================================================
%====================================================================

In order to include pulse control in both $\Lambda_\I$ and $\Lambda_\II$ systems we perform full quantum simulations for the second kind of CZ gate (see Table~\ref{table:one}). The fidelity of the operation is evaluated from the reduced density matrix of the two-qubit system before, $\rho(0)$, and after, $\rho=\rho(t_g)$, the gate. To reduce computational complexity we chose pulses for both $\Lambda_\I$ and $\Lambda_\II$ to be Gaussian-shaped and of the same bandwidth $\sigma$. We also allow a small overlap between pulses to reduce the total gate operation time [$\sigma(t_2-t_1)=\sigma(t_3-t_2)=3$]. The evolution is computed numerically using time-dependent Bloch-Redfield master equation
$i\dot{\rho} = [H+V(t),\rho]+\sum_s i\Gamma_s L_s\{\rho\}$
where 
$L_s=[P_s\rho P_s^\dag - (P^\dag_sP_s\rho+\rho P^\dag_s P_s)/2]$ \cite{nielsenchuang,MME,solenov1,solenov2,solenov2a,solenov2b}. We take the state $\rho(0)$ to be a pure state $\v{\psi_0}\bv{\psi_0}$. The fidelity of the gate operation is given by $F(\psi_0,\rho\{\psi_0\}) = (\bv{\psi_0}C(U)^\dag\,\rho\,C(U)\v{\psi_0})^{1/2}$. We compute the average fidelity integrated over all possible initial states of the two-qubit system \cite{fidel,fidel-com,super-operator}
\begin{eqnarray}\label{eq:avF}
F^2\!\! = \!\!\!\!\!\!\!\!\!\!\!\sum_{ijnm=\{1,4\}}
\!\!\!\!\!\!\!\!\!\frac{\delta_{in}\delta_{jm} + \delta_{ij}\delta_{nm}}{20} \bv{n}C(U)^\dag\rho\{\v{i}\bv{j}\}C(U)\v{m},
\end{eqnarray}
and optimize it over $\omega_C$. The resulting fidelity is shown in Fig.~\ref{fig3}(c) as a function of pulse bandwidth $\sigma$ and cavity quality factor $Q$. Decay of $^3E$ states of the NV center is dominated by the non-radiative recombination through states $^1A_1$ and $^1E$ ($\sim 10$ ns). The values for recombination rates were set to those reported in Ref.~\cite{Acosta}. In Fig.~\ref{fig3}(d) we show the fidelity as a function of pulse bandwidth, $\sigma$, and cavity frequency. The fidelity is the largest for values of $\omega_C$ corresponding to large values of $\de$ due to the interaction between the cavity mode and both defects. The maximum of the fidelity in Fig.~\ref{fig3}(d) is relatively broad, and, as a result, precise positioning of the defect and cavity energy levels with respect to each other is not necessary. The overall magnitude of fidelity is limited primarily by the cavity $Q$. The gates require the system to be in strong coupling regime, $\Delta/\gamma Q\ll 1$.

We have designed a framework for arbitrary all-optical two-qubit operations involving two NV centers coupled via a microcavity. This approach can also be applied to similar systems where fast all-optical two-qubit control is needed. The proposed framework can serve as the basis for development of efficient reversible and non-reversible \cite{Yale} two- and multi-qubit operations. By performing full quantum simulations of the system, we have demonstrated that  fidelity of the gates can reach high values for experimentally accessible parameters. 
Due to modest sensitivity of the fidelity to positioning of the cavity frequency,
the proposed approach can potentially be used to couple multiple defects pairwise using a single cavity mode without loss of fidelity due to their cross-talk. Thus, this cavity-based framework has an advantage of scalability in these systems.

D.S. gratefully acknowledges helpful discussion with S. G. Carter. This work was supported in part by the ONR, NRC/NRL, and by LPS/NSA. Computer resources were provided by the DOD HPCMP.

\end{document}